\documentstyle[preprint,aps]{revtex}
\begin{document}
\preprint{}
\title{
Finite temperature mobility of a particle coupled to a
fermion environment.}
\author{H. Castella\cite{adres}, X. Zotos}
\address{
 Institut Romand de Recherche Num\'erique en Physique des
Mat\'eriaux (IRRMA), \\
PHB-Ecublens, CH-1015 Lausanne, Switzerland}
\date{Received\ \ \ \ \ \ \ \ \ \ \ }
\bigskip\bigskip
\maketitle
\begin{abstract}
We study numerically the finite temperature and frequency mobility of a
particle coupled by a local interaction to a system of spinless fermions in
one dimension. We find that when the model is integrable (particle mass equal
to the mass of fermions) the static mobility diverges. Further, an enhanced
mobility is observed over a finite parameter range away from the integrable
point. We present a novel analysis of the finite temperature static mobility
based on a random matrix theory description of the many-body Hamiltonian.
\end{abstract}

\pacs{PACS numbers: 05.45.+b, 71.27.+a, 72.10.-d}

The description of dissipation in many-body systems
is a problem that traditionally attracted the interest of
theorists \cite{ford}. In particular, in classical systems,
a relation was observed between the occurence of normal heat transport
and chaotic behavior \cite{cas}.
Recently we have found a similar relation in the {\it finite temperature}
transport properties of a {\it quantum many-body} system \cite{cxp}.
This system, describing a particle coupled to a fermion
environment \cite{leg}, shows dissipationless transport at finite
temperatures when integrable while normal transport when non-integrable.

The analysis of the particle mobility was based on the
Kubo linear response theory \cite{Kubo};
within this framework, an indication of dissipationless transport is
given by a non-vanishing charge stiffness \cite{Kohn}
$D=\lim_{\omega\to 0} \omega\mu''(\omega)/2$, where $\mu''$ is the imaginary
part of the mobility. We have established that,
at any finite temperature $T>0$, $D$ remains non-zero when the model is
integrable, while $D$ decreases exponentially to zero with system size
in the generic non-integrable case \cite{xp}.
We attributed the non-vanishing $D$ to integrability motivated
by recent studies relating transport properties to
level statistics \cite{Wilk,mont} and level statistics to integrability
\cite{berry,Bell}. In particular, $D$ can be expressed as the
thermal average of level curvatures of the system
subject to an external flux and thus it reflects the properties of the energy
spectrum.

However integrability, and the resulting non-vanishing charge stiffness,
is a singular effect in parameter space. 
In this work, we are going to study if dissipationless transport is 
signaled by an enhanced static mobility in the neighborhood of the integrable 
point. In order to promote an effect as a signature of an integrable point, 
its robustness to small perturbations is indeed an important issue. 

The model we are studying, describes the evolution of a single particle on a
chain of $L$ sites with periodic boundary conditions. The particle is
coupled to an environment of $N$ spinless fermions through a local
repulsive interaction. The Hamiltonian is~:
\begin{equation}
\hat H=-t_h\sum_l (d^{\dagger}_{l+1} d_l +H.c.)
-t\sum_l (c^{\dagger}_{l+1} c_l + H.c. ) +
U\sum_l d^{\dagger}_l d_l c^{\dagger}_l c_l
\label{H}
\end{equation}
where $c_l(c^{\dagger}_l)$ are annihilation (creation) operators for the
fermions at site $l$ and $d_l(d^{\dagger}_l)$ for the particle. The model is
integrable by Bethe's ansatz for $t_h=t$ \cite{mcg} and non-integrable for
$t_h\neq t$; we have verified that the level-spacing distribution is Poisson
for the integrable case and follows the Gaussian Orthogonal Ensemble (GOE)
distribution for the non-integrable one.

From linear response theory, the real part of the mobility $\mu'(\omega)$
is~:
\begin{equation}
\mu'(\omega)=2\pi D\delta(\omega)+\pi{1-e^{-\beta\omega}\over \omega}
\sum_{\epsilon_n\neq \epsilon_m} p_n
\mid\langle n|\hat{\jmath}\mid m\rangle|^2
\delta(\omega-\epsilon_m+\epsilon_n).
\label{mu}
\end{equation}
where $\mid n\rangle$, $\epsilon_n$ denote the eigenstates and eigenvalues
of the Hamiltonian, $p_n=\exp(-\beta\epsilon_n)/Z$ the Boltzmann factors,
$\beta=1/T$, $Z$ the partition function and
$\hat{\jmath}=it_h\sum_l\left(d^{\dagger}_{l+1}d_{l}-
d^{\dagger}_{l}d_{l+1}\right)$ the current that does not commute with $\hat H$.
The f-sum rule relates the mobility to the average kinetic
energy $\hat{K}$ of the particle~: $\int_{-\infty}^{\infty}\mu(\omega)d\omega=
\pi\langle-\hat{K}\rangle$ where $\hat{K}=-t_h\sum_l (d^{\dagger}_{l+1}
d_l +H.c.)$ \cite{mald}.

Before dealing with the fine effect of integrability, we can comment on the
general frequency behavior of the mobility. For large
interaction parameter $U$, $\mu'(\omega)$ consists of two well separated
peaks at energies $\omega\simeq 0$ and $\omega\simeq U$. At high temperatures,
the weight and
the width of the high energy peak can be estimated from moments to be
$\pi\langle-\hat{K}\rangle\rho(1-\rho)$ and $\sqrt{2(3t_h^2+6t^2)}$
respectively \cite{thes};
the weight of the peak is essentially given by the number of particle-hole
pairs in the fermion bath while its width depends on the hopping matrix
elements. For smaller $U$, the two peaks merge into a single peak
at low frequencies.

Below we will present a study of
$\mu'(\omega)$ from a numerical simulation for finite systems. For the
smallest sytems, ($L\leq 12$), the energies and matrix elements of the
current are calculated exactly using a full-diagonalization technique.
For larger systems, ($L=14,16$), $\mu'(\omega)$ is evaluated
using a method proposed recently for finite temperature dynamical quantities,
combining random sampling of states with Lanczos diagonalization \cite{peter}.
In most cases, we will present results for the integrated mobility
$I(\omega)=\int_{-\omega}^{\omega}\mu'(\omega')d\omega'/(2\pi\langle-\hat{K}
\rangle)$, $\omega>0$, thus avoiding the usual artificial broadening of the
$\delta$-functions in Eq.(\ref{mu}).
$I(\omega)$ has two simple limits, $\lim_{\omega\to 0}I(\omega)=D/
\langle-\hat{K}\rangle$ and $\lim_{\omega\to\infty}I(\omega)=1/2$. The slope
of $I(\omega)$ gives the mobility itself.

Our investigation is restricted to the high
temperature regime, to relatively small $U=2t$ and at half filling,
$\rho=N/L=1/2$. The high temperature
regime is indeed easier to analyse numerically since the finite size
effects are less pronounced and the choice of $U$ allows to maximize the
effect of integrability since the charge stiffness decreases with increasing
$U$. In the following we present numerical results for
$\mu'(\omega)$, discuss the static mobility, $\mu_0=\lim_{\omega\to 0}
\mu'(\omega)$ around the integrable point and estimate 
the finite size effects. We then propose a new approach for 
evaluating $\mu_0$ based on a recent relation between
dissipation and random matrix theory (RMT) \cite{Wilk,mont}. We conclude with
a presentation of $\mu_0$ as a function of $t_h$.

In Fig. \ref{one} we present $\mu'(\omega)$ for different size systems
at $T=5t$ and far from the integrable point. We note that $\mu'(\omega)$ is
well defined with no excessive size dependence (the restricted number of
Lanczos iterations in the approximate method
does not allow us to reach the lowest frequencies for $L=14,16$).
The size dependence is a very
important issue since normal transport can be established only if
$\mu_0$ is intensive, i.e. if it saturates with system size \cite{cas}.
The range of available sizes is rather limited but nevertheless we 
observe that $\mu_0$ has mostly saturated for $L=12$.
Therefore we can expect to have real transport in the high temperature
regime even for such small systems. The indicated full points at $\omega=0$,
in this and the following figures, are
the estimation of $\mu_0$ by the RMT approach that will be discussed below.

Fig. \ref{two} presents $I(\omega)$ for $L=12$ and $T=5t$ scanning $t_h$
through the integrable point $t_h=t$. We first note that at high frequencies,
$\omega>1.5t$, the particularity of the integrable point is not reflected
in $I(\omega)$. On the other hand, we observe that, at low frequencies,
all the weight is concentrated in the charge stiffness for $t_h=t$,
while it spreads over a finite frequency range for $t_h\neq t$.

We now focus on the static mobility,
around $t_h=t$. It is a very difficult quantity to extract from
numerical results on finite systems because of the discreteness of
the spectrum. We found convenient to
estimate $\mu_0$ from the $\omega\to 0$ limit of $(I(\omega)-I(0))/\omega$
which is presented in Fig. \ref{three} for different $t_h$.
We note that the low frequency limit is well defined for $t_h\geq 1.5t$; 
it becomes more difficult to extract $\mu_0$ for $t_h\simeq t$ as 
$(I(\omega)-I(0))/\omega$ varies rapidly close to $\omega=0$. Nevertheless, 
we see that the extrapolated $(I(\omega)-I(0))/\omega$ gives a value of 
$\mu_0$ that clearly {\it increases} as $t_h$ approaches the integrable 
point $t_h=t$.

In contrast, we would like to illustrate the low
frequency behaviour of $\mu'(\omega)$ for $t_h=t$.
Fig. \ref{four} shows $I(\omega)$
for different system sizes up to $L=16$. The stars at $\omega=0$ indicate
the exact charge stiffness in the thermodynamic limit as was
computed analytically  in a previous work using the Bethe-ansatz method
\cite{cxp}. The weight
at low frequencies is concentrated in the charge stiffness $I(\omega=0)$ and
$I(\omega)$ reaches this point with a zero slope indicative of a $\mu_0=0$.
Although we still observe a residual size dependence, we think that the
depletion of weight at low frequencies persists in the thermodynamic limit
since $I(\omega=0)$ is already close to the analytical value.
When $U$ is decreased the shoulder at $\omega\simeq t$ moves to lower
frequencies
and the depletion of weight is observed in a smaller frequency range.
These observations suggest that the effect of integrability shows
up in the low frequency range, i.e. for $\omega$ lower than all the
natural energy scales, $t,U$. 

From a first point of view we would expect 
the effect of integrability, absence of level repulsion and modification of 
level density - density correlations, to be reflected only in a frequency 
range of the order of the level spacing $\Delta$ (in which case it might be 
amenable to a RMT analysis as in e.g. \cite{Tani}). However, in our many-body 
system, $\Delta$ scales to an exponentially small value with the size of 
the system (of the order of $10^{-2}t$ for $L=12$). The reason that our 
results for $\mu'(\omega)$ show the influence of integrability at 
$\omega \gg \Delta$, is the strong dependence of the current 
matrix elements on $|t_h-t|$ in this frequency range. Nevertheless, 
as we will show in the next section, we can still obtain an estimate of  
$\mu_0$ by assuming that, {\it locally} in the spectrum, the current matrix is 
described by a random matrix. 

{\it $\mu_0$ from random matrix theory} : In the non-integrable case,
we expect the Hamiltonian matrix of our system to be described, at least 
locally, by random matrices from the GOE ensemble (as we verified for the 
level- spacing distribution). We can then rewrite the Kubo formula for the 
static mobility $\mu_0$ using a relation between
off-diagonal and diagonal elements of the current for random matrices in
this ensemble \cite{Wilk}~:
\begin{equation}
2~\langle\mid j_{n m}\mid^2\rangle_{\epsilon_n\neq\epsilon_m}=
\langle\left(j_{n n}-\bar{v}(E)\right)^2\rangle=\sigma^2(E)
\label{wi}
\end{equation}
Here $j_{n m}=\langle n\mid\hat{\jmath}\mid m\rangle$,
$\bar{v}=\langle j_{n n}\rangle$ and the outer brackets
denote an average of matrix elements locally in the spectrum, i.e. for
$\epsilon_n,\epsilon_m\simeq E$ (within a given translational symmetry
sector). Numerical tests show that indeed, for
$t_h\neq t$, the above relation (\ref{wi}) is well satisfied.
From (\ref{mu}) the static mobility $\mu_0$
can be expressed in terms of the variance $\sigma^2$ and the
many-body density of states $\rho(E)$ at the energy $E$~:
\begin{equation}
\mu_0={\beta\pi\over 2}\int_{-\infty}^{\infty} {\exp(-\beta E)\over Z}
\rho^2(E)\sigma^2(E)dE.
\label{mud}
\end{equation}
(a sum over all translational symmetry sectors is implied). 
This expression may be obtained from $(I(\delta\omega)-I(0))/
\delta\omega$, for $\delta\omega\gg\Delta$ but much smaller than the 
natural energy scales, which is the appropriate 
limit for finding the static response in a macroscopic system. 

Further, as the diagonal elements of the current can be calculated  through
the first derivatives of the energies with respect to a fictitious flux $\phi$,
$j_{n n}=\partial \epsilon_n/\partial\phi$, we see that (\ref{mud})
relates a transport property, $\mu_0$, to a change in boundary conditions;
$\phi$ is introduced in the Hamiltonian by the usual Peierls construction
$t_h\to t_he^{i\phi}$.  Similar relations have been used in the
theory of localization \cite{Thoul} and in the study of the conductivity of
metallic systems at zero temperature \cite{mont} (see also \cite{faas}).
We propose it here for a {\it many-body} system at {\it finite temperature} .

We use expression (\ref{mud}) to evaluate the static mobility
for our model in the high temperature regime,
$\beta\to 0$. Our finite systems do not allow a test of this expression
at low temperatures because there might appear deviations from
GOE statistics in the low energy part of the spectrum. The $\mu_0$ so
extracted was shown in Fig. \ref{one} and Fig. \ref{three} by the
full points at $\omega=0$ in fair agreement with
the extrapolated value from $\mu'(\omega)$. These results give support to
the RMT approach to transport for our model.
In Fig. \ref{five} we collect our results for $\mu_0$ as a function of $t_h$
for different size systems. For a fixed size, we observe a divergence
of $\mu_0$ at $t_h=t$ which becomes more {\it pronounced} 
with {\it increasing size}. This indicates that in the infinite size limit 
an enhanced $\mu_0$ appears over a finite parameter range around the
integrable point. Further study of the critical behaviour at $t_h=t$
would require larger size systems.

In consistence with a finite $\mu_0$, the above analysis can also
give some insight into the suppression of the charge stiffness
for our system as follows~:
the second derivative of the free energy, $F$, with respect to
the flux $\phi$~ is~:
\begin{equation}
{\partial^2 F\over\partial\phi^2}=2D-\beta\sum_n p_n\left({\partial\epsilon_n
\over\partial\phi}\right)^2+\left(\sum_n p_n{\partial\epsilon_n\over\partial
\phi}\right)^2
\end{equation}
In this expression the third term vanishes by symmetry and we find that
in a system as ours, without persistent currents,
${\partial^2 F\over\partial\phi^2}$ vanishes in the thermodynamic limit.
Therefore the charge stiffness is also given by~:
\begin{equation}
D={\beta\over 2}\sum_n p_n \left({\partial\epsilon_n\over\partial\phi}\right)^2
={\beta\over 2}\int_{-\infty}^{\infty} {e^{-\beta E}\over Z} \rho(E)\left(
\sigma^2(E)+\bar{v}^2(E)\right)dE
\end{equation}
If the current matrix can be described by a random matrix locally in the 
spectrum, the variance $\sigma^2$ has to scale with system size
as $1/\rho(E)$; this is consistent with a finite $\mu_0$ 
for the thermodynamic limit in Eq.(\ref{mud}). Moreover we observed that 
$\bar{v}$ is 
microscopic in the bulk of the spectrum. Therefore, the charge stiffness 
itself is exponentially suppressed in the thermodynamic limit system since
$\rho(E)\propto\exp(\alpha L)$. It would be interesting to see
whether the RMT approach for the finite temperature static conductivity also 
applies to other generic non-integrable systems in the metallic regime.

In conclusion, the numerical simulations presented in this
work allowed us to establish a scenario for the signature of
integrability in the finite frequency mobility. In general, the
dependence of $\mu'(\omega)$ is governed by $U$, $t_h$ and
$t$; at low frequencies, the weight is concentrated in a peak centered at zero
frequency of roughly constant magnitude but of decreasing width as the
integrable point is approached. For $t_h=t$ this peak collapses to a
$\delta$-function with weight equal to the charge stiffness.
As the parameter range over which an enhanced static mobility is observed is
finite, this study suggests that the effect of integrability on transport is a
physically relevant effect.

\acknowledgments
We would like to thank P. Prelov\v sek and J. Jakli\v c for the help with the
Lanczos method and useful discussions.
This work was supported by the Swiss National Fund Grant No.
20-39528.93, the University of Geneva and the University of Fribourg.

\begin{figure}

\caption{Normalized finite frequency mobility for $t_h=2t$, $U=2t$, $T=5t$ 
and different system sizes. The full point at $\omega=0$ indicate the static
mobility from the RMT approach for $L=14$ and $\beta\to 0$.}
\label{one}

\caption{Integrated mobility $I(\omega)$ for different $t_h$ at $U=2t$,
$T=5t$ and $L=12$ : $t_h=t$ (thick solid line), $0.5t$ (solid line), $0.7t$
(dotted line), $1.2t$ ( dashed line ) and $2t$ (long dashed line).}
\label{two}

\caption{$(I(\omega)-I(0))/\omega$ as a function of frequencies for $L=12$,
$U=2t$, $T=5t$ and several non-integrable cases in the low frequency regime. 
The full points at $\omega=0$ indicate the static mobility from the
RMT approach.}
\label{three}

\caption{Integrated mobility $I(\omega)$ for the integrable case $t_h=t$ at
$U=2t$ and different temperatures and sizes : $L=16$ ( solid line ),
$L=14$ ( dotted line ), $L=12$ ( dashed line ) and $L=10$ ( long dashed
line ). The stars indicate the exact charge stiffness in the thermodynamic
limit.}
\label{four}

\caption{Normalized static mobility as a function of $t_h/t$ in the
high temperature limit and for $U=2t$. The values are computed within
the RMT approach.}
\label{five}

\end{figure}


\begin{references}
\bibitem[*]{adres}  Present address : Department of Physics, Ohio State 
University, 174 West 18th Ave., Columbus, OH, 43210-1106.
\bibitem{ford} J. Ford, Physics Rep. {\bf 213}, 271 (1992).
\bibitem{cas} G. Casati, J. Ford, F.V. Vivaldi, and W.M. Visscher, Phys.
Rev. Lett. {\bf 52}, 1861 (1984); E. A. Jackson and A.D. Mistriotis,
J. Phys. : Condens. Matter {\bf 1}, 1223 (1989).
\bibitem{cxp} H. Castella, X. Zotos, and P. Prelov\v sek, Phys. Rev. Lett.
{\bf 74}, 972 (1995).
\bibitem{leg} A.O Caldeira and A.J. Legett, Phys. Rev. Lett. {\bf 46},
211 (1981);
A.H. Castro Neto and A.O. Caldeira, Phys. Rev. Lett. {\bf 67}, 1960 (1991).
\bibitem{Kubo} R. Kubo, J. Phys. Soc. Jpn., {\bf 12}, 570 (1957).
\bibitem{Kohn} W. Kohn, Phys. Rev. {\bf 133},171 (1964).
\bibitem{xp} a similar effect has been studied in a system of interacting
spinless fermions : X. Zotos and P. Prelov\v sek, Phys. Rev. B (1996).
\bibitem{Wilk} M. Wilkinson, J. Phys. A : Math. Gen. {\bf 21}, 4021 (1988);
Phys. Rev. A {\bf 41}, 4645 (1990).
\bibitem{mont} E. Ackermans and G. Montambaux Phys. Rev. Lett. {\bf 68}, 642
(1992); B.D. Simons, A. Szafer, and B.L. Altshuler, JETP Lett. {\bf 57},
276 (1993); B. D. Simons and B.L. Altshuler, Phys. Rev. Lett. {\bf 70}, 4063
(1993).
\bibitem{berry} M.V. Berry and M. Tabor, Proc. Roy. Soc. London, Ser. A
{\bf 356}, 375 (1977).
\bibitem{Bell} D. Poilblanc, T. Ziman, J. Bellisard, F. Mila, and G.
Montambaux, Europhys. Lett. {\bf 22}, 537 (1993).
\bibitem{mcg} J.B. McGuire, J. Math. Phys. {\bf 6}, 432 (1965);
J. Math. Phys. {\bf 7}, 123 (1966); E.H. Lieb and F.Y. Wu, Phys. Rev. Lett.
{\bf 20}, 1445 (1968); X. Zotos and F. Pelzer, Phys. Rev. B {\bf 37}, 5045
(1988).
\bibitem{mald} P.F. Maldague, Phys. Rev. B {\bf 16}, 2437 (1977);
D. Baeriswyl, C. Gros, and T.M. Rice, Phys. Rev. B {\bf 35}, 8391 (1987).
\bibitem{thes} H. Castella, Ph.D. thesis, University of Geneva (1995).
\bibitem{peter} J. Jakli\v c and P. Prelov\v sek Phys. Rev. B
{\bf 49}, 5065 (1994). 
\bibitem{Tani} for a study of the $\omega\simeq\Delta$ regime, see N. 
Taniguchi, B. Altshuler, Phys. Rev. Lett. {\bf 71}, 4031 (1993).
\bibitem{Thoul} J.T. Edwards and D.J. Thouless, J. Phys. C: Solid St. Phys.
{\bf 5}, 807 (1972); D.J. Thouless, Phys. Rev. Lett. {\bf 39}, 1167 (1977).
\bibitem{faas} M. Faas, B.D. Simons, X. Zotos, and B.L. Altshuler, Phys.
Rev. A {\bf 48}, 5439 (1993).
\end{references}
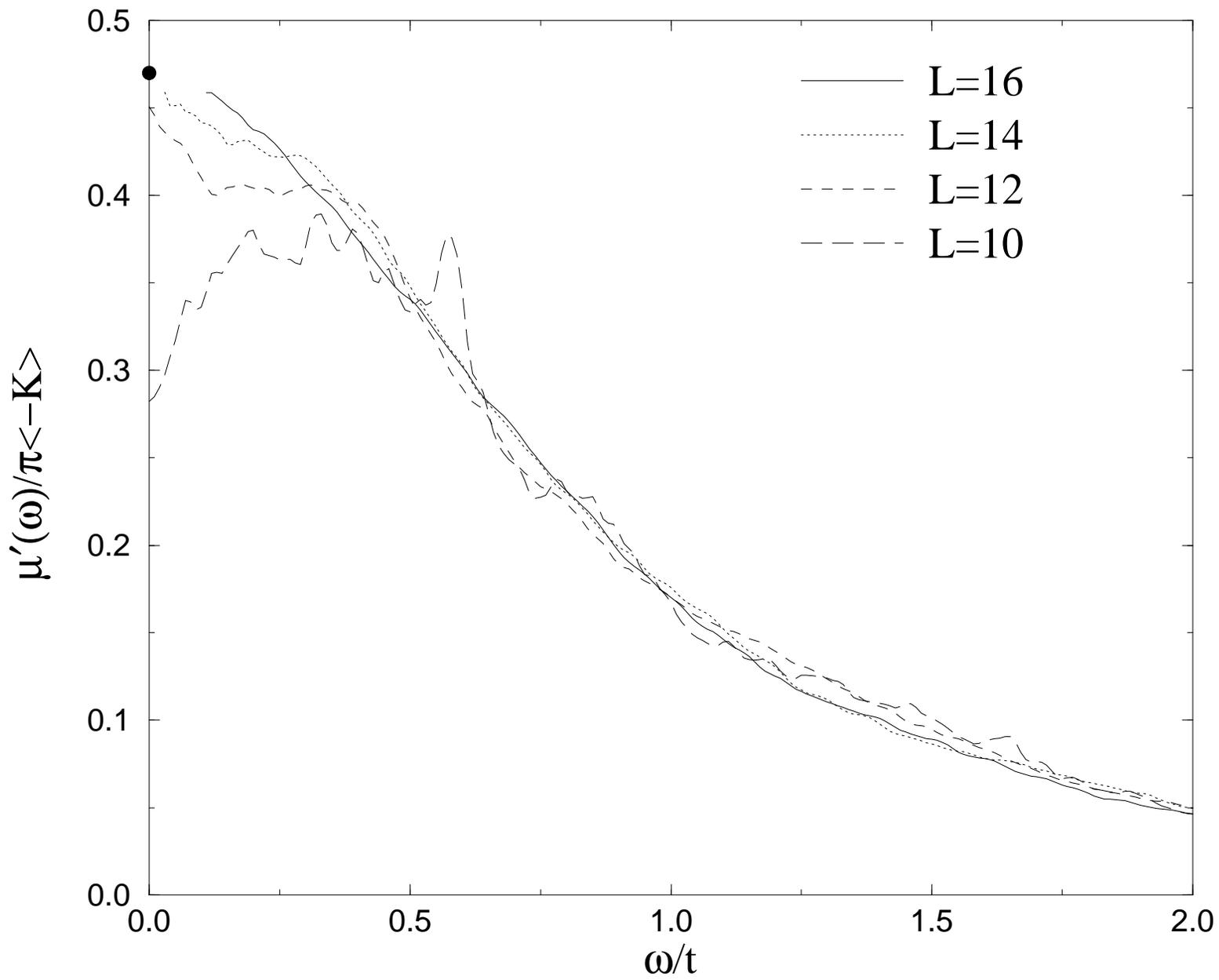
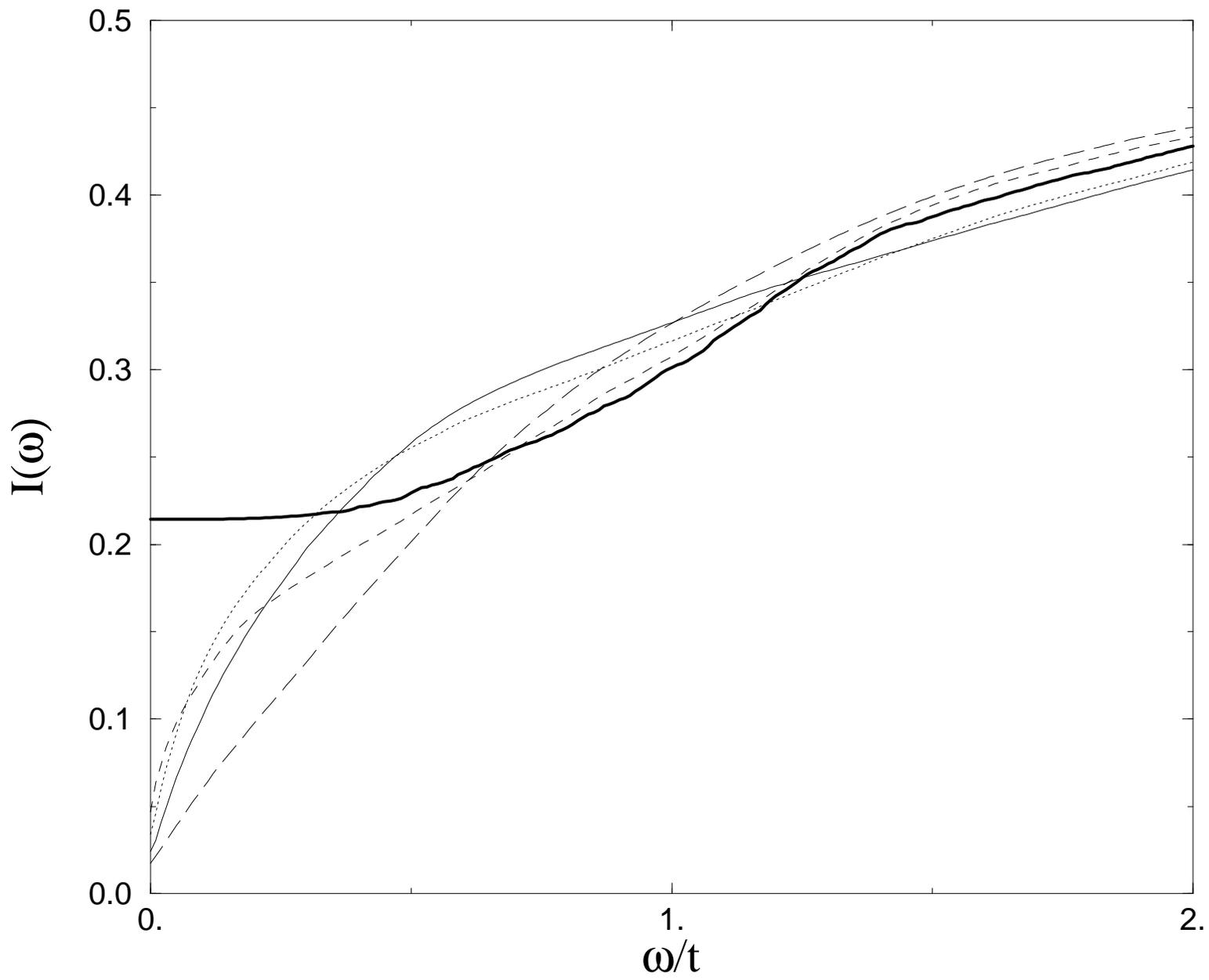
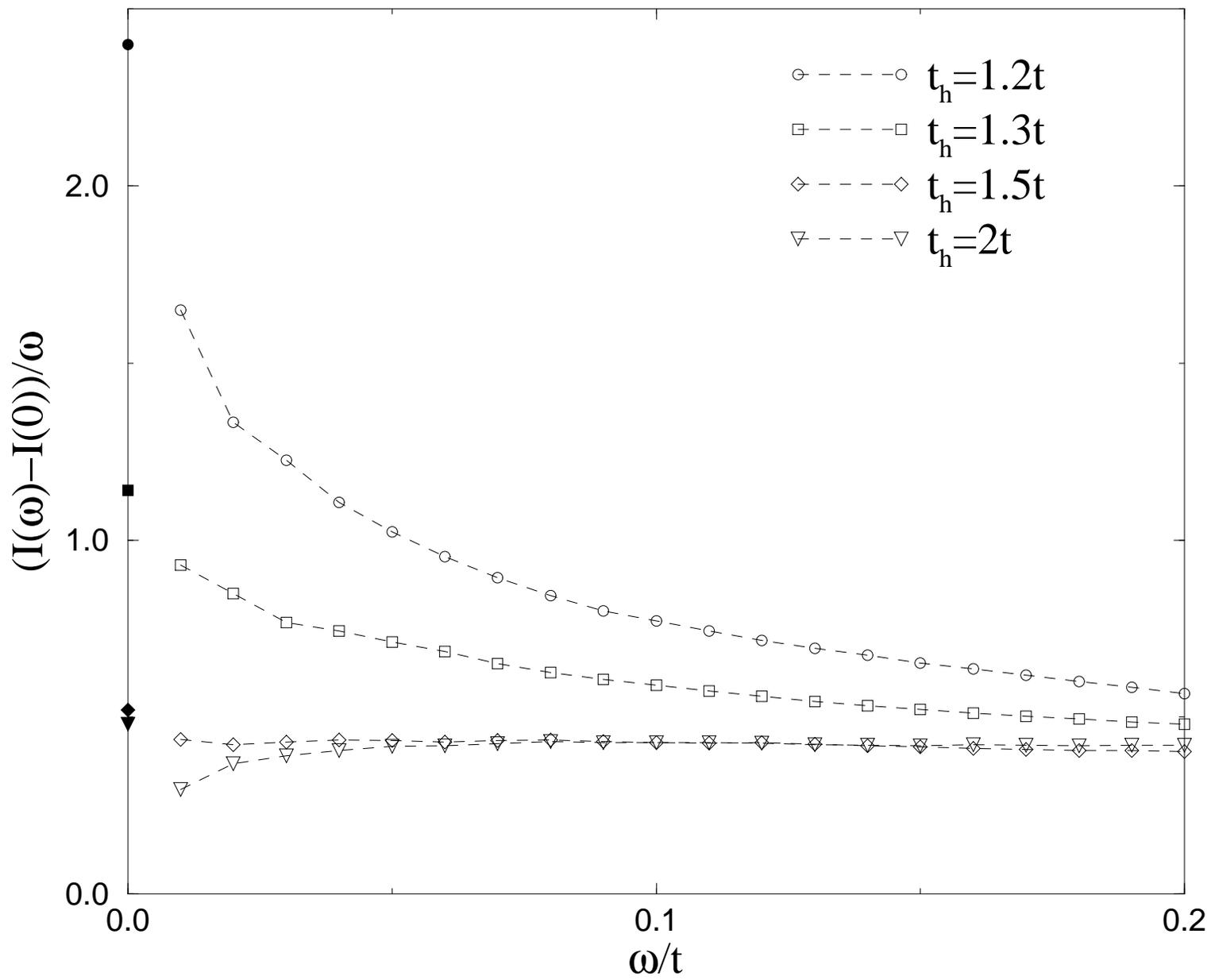
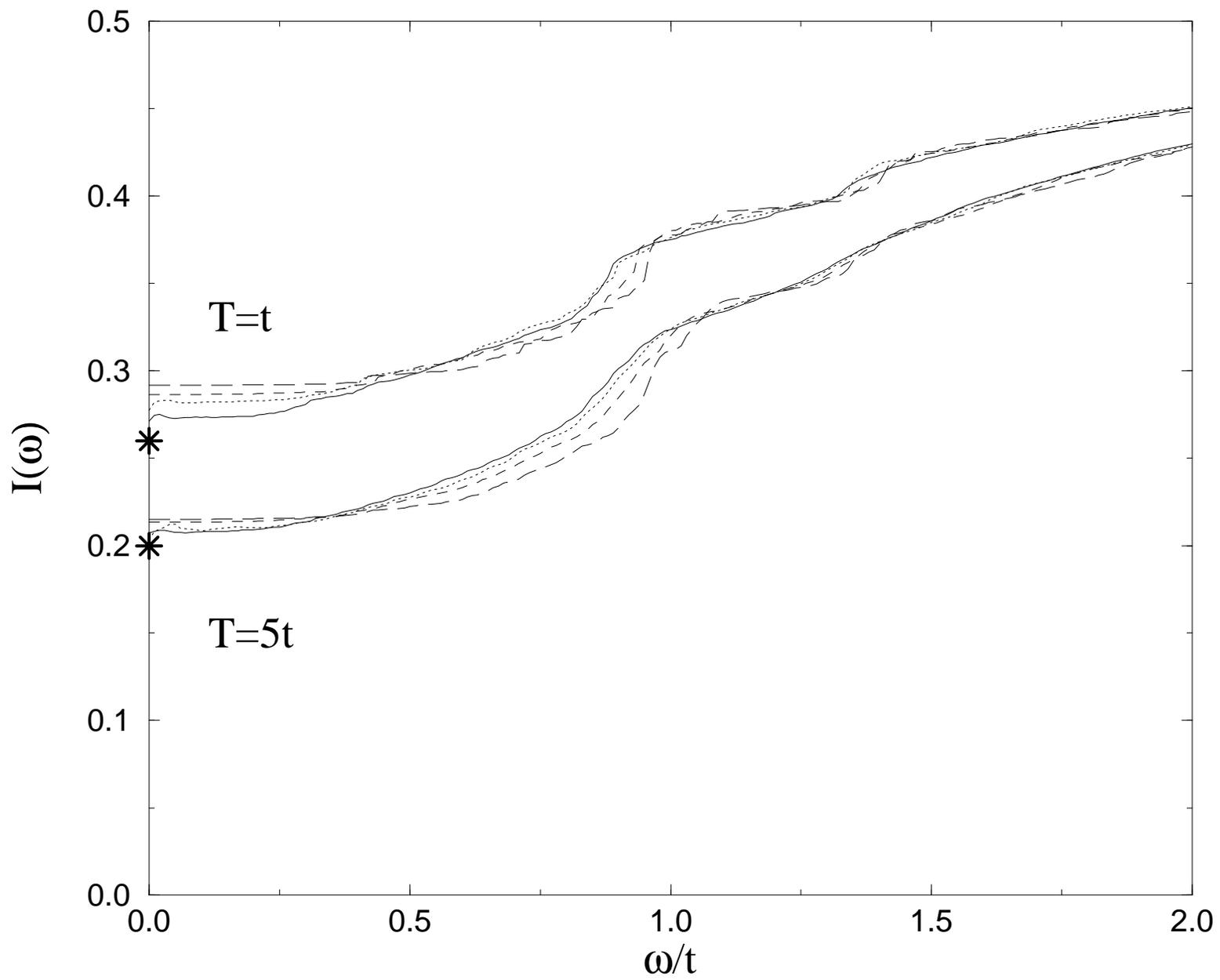
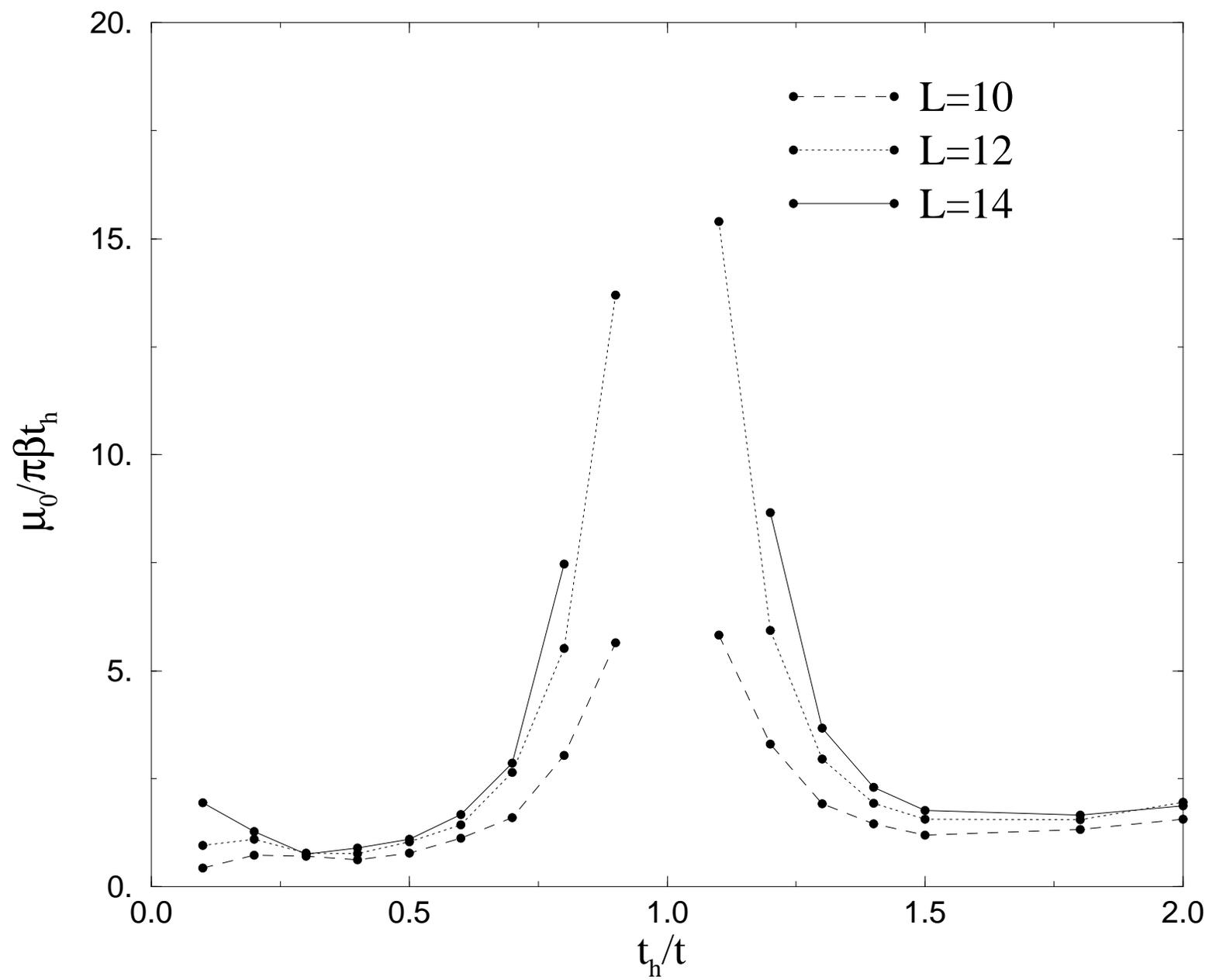
\end{document}